# Why Do Thick MOCVD-Grown β-$Ga_2O_3$ Epilayers on (001) Substrates Crack: Crystallographic Origin

Martin Frentrup[1,*], Indraneel Sanyal[2,*], Dan Lamb[3], Ciaran P. Llewelyn[3], Jonathan Evans[3], Owen J. Guy[3,4], Mike Jennings[3,5], Saptarsi Ghosh[3,5,#]

[1]Department of Materials Science and Metallurgy, University of Cambridge, Cambridge, CB3 0FS, UK
[2]AIXTRON Ltd, Buckingway Business Park, Anderson Rd, Swavesey, Cambridge CB24 4FQ, UK
[3]Centre for Integrative Semiconductor Materials, Swansea University, Swansea SA1 8EN, UK
[4]Department of Chemistry, Swansea University, SA1 8EN, UK
[5]Department of Electronic and Electrical Engineering, Swansea University, SA1 8EN, UK

*These authors contributed equally.

# Author to whom correspondence should be addressed: saptarsi.ghosh@swansea.ac.uk

**Abstract**

Thick, defect-free epitaxial layers grown using industry-standard techniques are a fundamental requirement for the widespread adoption of fully vertical power devices based on ultra wide-bandgap gallium oxide ($Ga_2O_3$). However, metal–organic chemical vapour deposition (MOCVD) of such layers on native β-$Ga_2O_3$ substrates with the largest-diameter (001) orientation remains relatively unexplored, and the origins of the reported surface roughening and cracking with increasing thickness are not yet fully understood. To address this, we report a systematic study of MOCVD-grown β-$Ga_2O_3$ epilayers deposited at growth rates of ~3.5 μm $h^{-1}$, with thicknesses from 0.3 to 3.5 μm. The epilayers exhibit a relatively smooth but striated surface morphology, with progressively increasing nanometre-scale roughness beyond coalescence and crack formation observed from ~1.8 μm thickness. High-resolution X-ray diffraction reveals that, despite growth on (001) substrates, the epilayers adopt a predominantly (−401)-oriented structure from the earliest stages of growth. Rocking curve analysis further indicates a higher degree of in-plane twist than tilt, both decreasing with increasing epilayer thickness. While the epilayer and substrate are lattice-matched along the [010] in-plane direction, the epitaxial alignment in the orthogonal epilayer [104] in-plane direction imposes, in theory, a maximum tensile in-plane strain of approximately +4.1% arising from the underlying lattice mismatch, thereby driving crack formation perpendicular to this direction. Our results suggest that this epitaxial relationship is likely associated with faceted reconstruction of the (001) substrate surface during annealing, driven by the minimisation of surface energy under oxygen-rich MOCVD growth conditions.

## Introduction

Ultra-wide-bandgap semiconductors such as diamond, AlN, and $Ga_2O_3$ have recently emerged as the next frontier for more efficient power electronics to meet the growing energy demands of modern society. Among these, β-phase $Ga_2O_3$, with an estimated critical electric field for breakdown of ~8 MV $cm^{-1}$ and controllable n-type doping ranging from $10^{15}$ $cm^{-3}$ to $10^{20}$ $cm^{-3}$, is particularly promising for unipolar power devices[1,2]. Uniquely among wide- and ultra-wide-bandgap semiconductors, low-cost native β-$Ga_2O_3$ substrates with very low background impurity concentrations and intrinsic defect densities can be grown from the melt[3]. Such 100 mm (4-inch) diameter substrates are already commercially available, with a clear roadmap toward 150 mm (6-inch) substrates[4]. The availability of native substrates offers a strong industrial advantage for mass production of fully vertical $Ga_2O_3$ transistors, where current flows between top source and bottom drain electrodes, simplifying electric-field and thermal management. This design also enables breakdown-voltage scaling with drift-layer thickness while independently tailoring cell pitch and device area to the ON-current requirement, analogous to practices in traditional silicon and SiC devices.

Among melt-grown methods, edge-defined film-fed growth (EFG) remains the most scalable technique for β-$Ga_2O_3$ substrates, overcoming the thermal challenges associated with the Czochralski method[4]. In EFG, β-$Ga_2O_3$ crystals are commonly pulled along [010][5] while adopting a (001) principal surface, such that the (001) cleavage plane is not critically stressed during growth. Consequently, (001)-oriented β-$Ga_2O_3$ substrates are available in the largest sizes and at the lowest areal cost. On such (001)-oriented substrates, halide vapour phase epitaxy (HVPE)-grown, low-doped (001) epilayers enabled the realisation of the first high-voltage vertical gallium oxide Schottky barrier diode (SBD)[6]. Since then, using different termination strategies and dielectrics, several groups have made steady progress in device development on the (001) orientation, achieving multi-kilovolt breakdown voltages.[7,8,9,10]

However, while HVPE currently remains the preferred approach for homoepitaxy of (001) β-$Ga_2O_3$ — where homoepitaxy conventionally implies crystallographic continuity between epilayer and substrate — owing to its relatively high growth rates (~5–15 μm $h^{-1}$)[11], it nevertheless exhibits inherent technological limitations that hinder industrial adoption. For instance, polycrystalline defects, stacking faults, and particulates generated during HVPE growth are known to cause device-to-device variability[12,13]. In addition, as-grown epilayers often exhibit non-ideal surface morphologies[14–16], necessitating chemical–mechanical polishing (CMP) before device fabrication. This additional step may remove as much as ~30–50 % of the as-grown material[6,17], increasing cost and waste, potentially introducing impurities, and severely constraining heterojunction design owing to the substantial and uncertain material removal in polishing.

In contrast, metal–organic chemical vapour deposition (MOCVD) is a non-equilibrium growth technique whose multi-wafer industrial scalability is well established. It is the preferred epitaxial method for complex III–nitride and III–arsenide heterostructures used in photonics and electronics and could overcome several limitations inherent to HVPE. Nevertheless, as noted in a recent review by Xu *et al.*, investigations of homoepitaxy on (001) gallium oxide native substrates using MOCVD remain limited[11]. Among the reported studies, most focus on relatively thin epilayers (~0.2 µm to 1.0 µm) and primarily on surface roughness[18,19,20,21]. From a device standpoint, increased surface roughness can lead to inhomogeneous adhesion of metals and dielectrics, resulting in degraded electron transport and non-uniform device performance across the wafer.

For device-relevant thicker layers, a more critical challenge arises: crack formation. This has been reported by Meng and coworkers[22], who achieved higher growth rates (~3 µm $h^{-1}$) by using trimethylgallium (TMGa) instead of triethylgallium (TEGa); however, the resulting thicker epilayers exhibited severe cracking. In subsequent work[23], various $(Al_xGa_{1-x})_2O_3$ buffers were introduced, but these were not fully successful in mitigating crack formation. Cracks are known to induce premature device breakdown by producing localised electric-field peaks. Although detailed morphological information was not reported, similar cracking could reasonably be expected to exacerbate breakdown degradation in ~3.5 µm thick epilayers grown on (001) substrates relative to those co-grown on (010) substrates by Sundaram et al.[24]. Beyond such performance impacts, cracking is inherently detrimental to power-device fabrication, significantly reducing yield and reliability. Therefore, understanding the origin and evolution of these growth non-idealities — namely surface roughness and cracking as a function of epilayer thickness in MOCVD-grown films on (001) substrates is a pressing requirement for the near-term commercialisation of gallium oxide technology.

To address this gap in understanding, we present a systematic investigation of a series of $Ga_2O_3$ epilayers grown by MOCVD in an industry-standard large-area cold-wall reactor. By examining the evolution of surface topography and microstructure with increasing epilayer thickness, we identify the physical origins of the observed morphological and structural features. While cracking and anomalous diffraction features have been reported previously, this work establishes the underlying crystallographic relationship between the epilayer and substrate and provides a strain-driven framework linking anisotropic lattice mismatch to crack formation. These findings identify this epitaxial relationship as the key limitation to achieving high-quality thick layers on (001) bulk substrates under conventional growth conditions, while suggesting pathways for overcoming it.

**Experiments**

Prior to epitaxial growth of device-quality layers, thermal annealing of substrates inside the reactor is routinely employed for many semiconductor systems, including gallium oxide[19]. Accordingly, for each growth run, the substrate temperature was first ramped to 850 °C under oxygen, and the (001) substrate was annealed at this temperature under a pressure of 100 mbar with an oxygen flow rate of 2000 sccm (~89.3 mmol/min) for 10 min. Epilayer growth followed immediately after annealing, using the same oxygen flow rate but at a reduced temperature of 825 °C and a chamber pressure of 50 mbar. During growth, the TMGa flow rate was maintained at 176 µmol $min^{-1}$, corresponding to a VI/III ratio of ~500. Under these conditions, a growth rate of ~3.5 µm $h^{-1}$ was achieved, representing a comparatively high rate for MOCVD growth of $Ga_2O_3$ while maintaining industry-standard scalability. The investigated series comprised five epilayers (samples B–F) with thicknesses of 0.3 µm (B), 1.2 µm (C), 1.8 µm (D), 2.3 µm (E), and 3.5 µm (F), grown under identical conditions except for the growth duration, which varied from 5 min to 1 h. A control sample (sample A), which was cooled immediately after annealing without undergoing growth, was also included for comparison.

### Results and Discussion

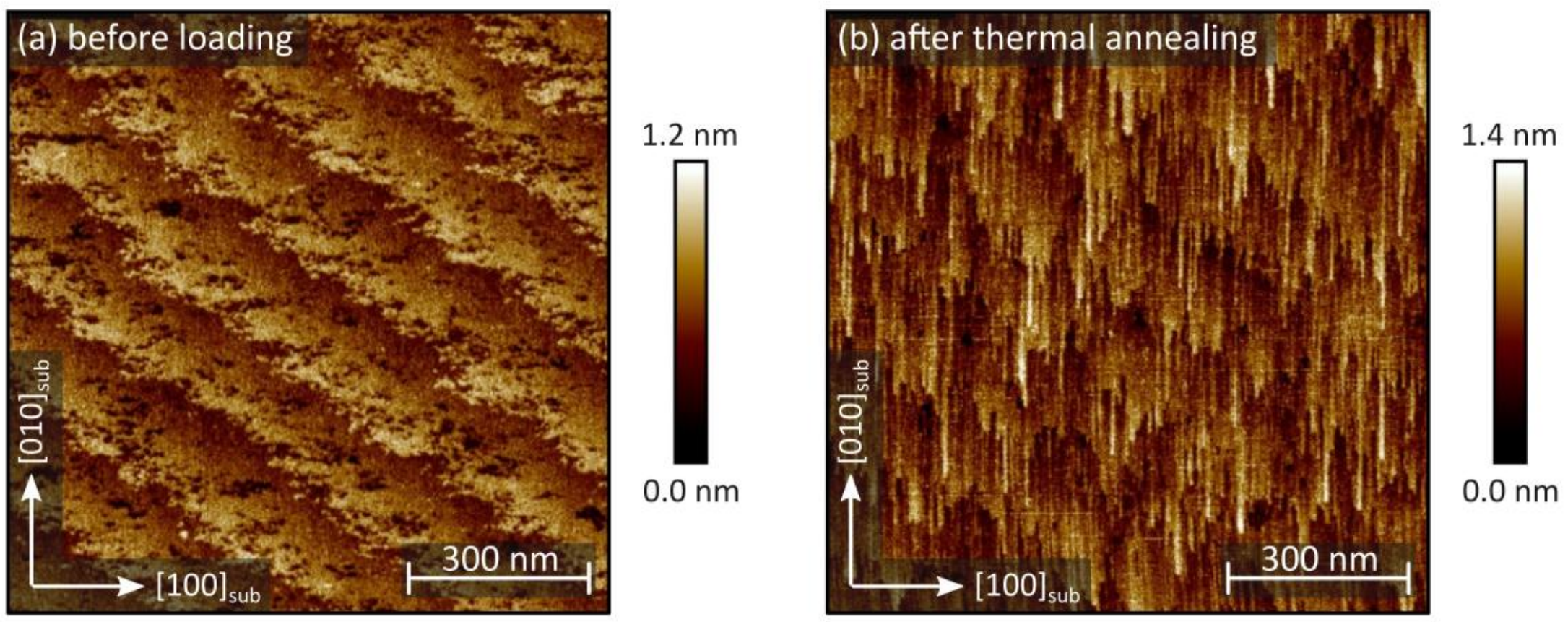

*Figure 1: Representative 1×1 μm² AFM images of the (001) β-$Ga_2O_3$ substrate morphology (a) before loading into the reactor and (b) after the thermal annealing step inside the reactor.*

Figure 1(a) shows the surface topography of the substrate before reactor loading. The surface consists of flat terraces separated by ~0.6 nm steps, corresponding to a full unit-cell step normal to the (001) plane of monoclinic β-$Ga_2O_3$. The comparatively large terrace widths (~150 nm) are consistent with the nominal effective miscut (~0.2°) from the low-index (001) plane. However, as shown in Figure 1(b), this topography is not fully preserved in the annealed control sample (sample A). First, the terraces are no longer atomically flat but instead consist of periodic facets. Second, although the terrace orientations are mostly retained, the step edges become ill-defined as the facets elongate to different extents, uniformly along the [010] substrate direction. Importantly, the comparable height scales in Figures 1(a) and 1(b) confirm the absence of step bunching.

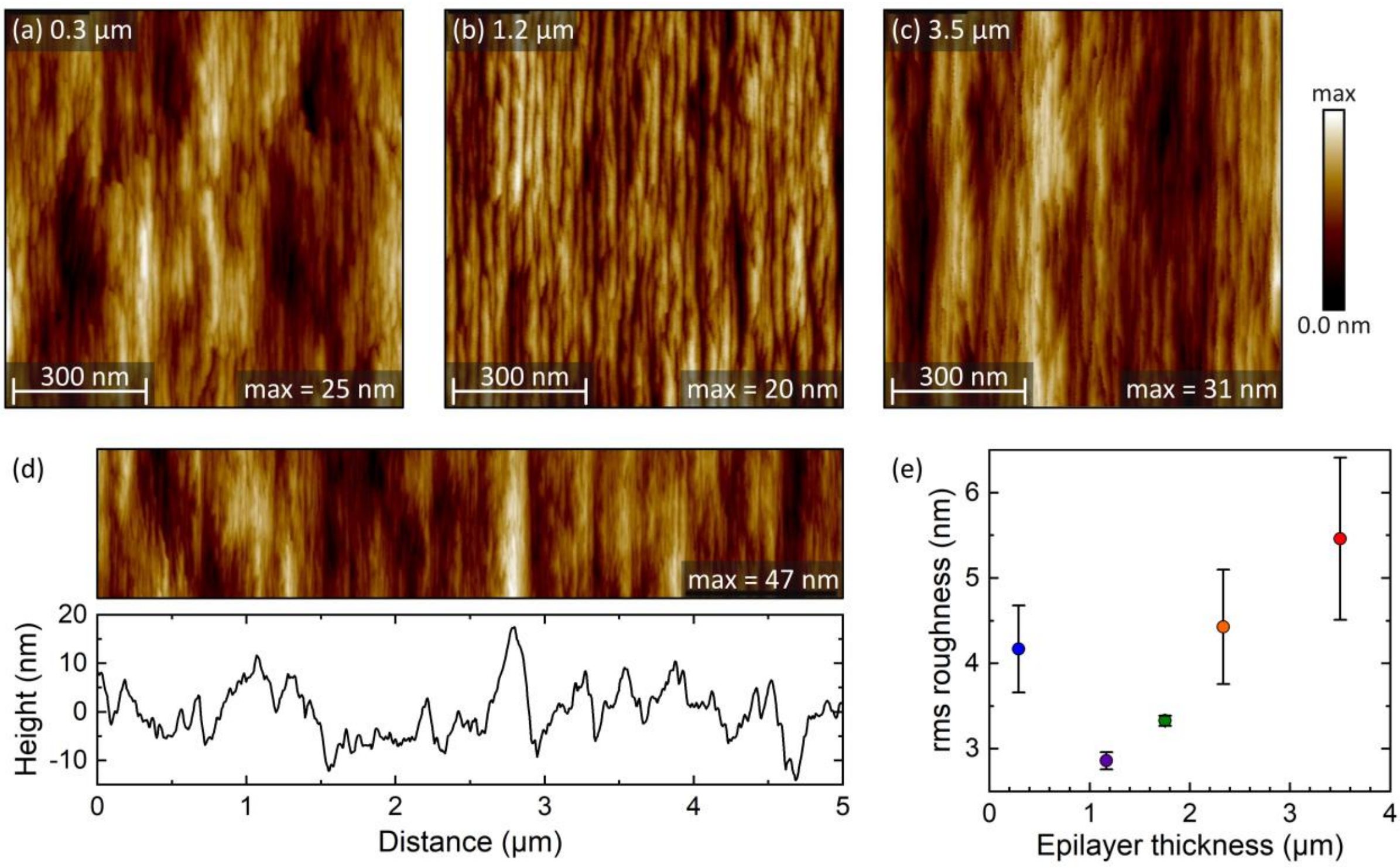

*Figure 2: Representative 1×1 µm² AFM images of β-$Ga_2O_3$ epilayers with thicknesses of (a) 0.3 µm, (b) 1.2 µm, and (c) 3.5 µm grown on the (001) substrates. The striations observed in the epilayers are aligned with the [010] direction of the underlying (001) substrate. (d) A wider-area 5×1 µm² AFM image of the 3.5 µm thick epilayer, together with the corresponding line profile averaged across the full image width. (e) Evolution of the RMS roughness as a function of epilayer thickness, extracted from multiple area scans.*

Post-growth evolution of the surface topography as a function of epilayer thickness is shown in Figure 2. At an epilayer thickness of 0.3 µm (sample B; Figure 2(a)), the surface is nearly coalesced and is dominated by dense striations ~500 nm in length. These striations are oriented parallel to the [010] direction of the underlying substrate and align with the elongated features observed in Figure 1(b). With increasing epilayer thickness, this striated morphology persists, as illustrated for samples with thicknesses of 1.2 µm (sample C) and 3.5 µm (sample F) in Figures 2(b) and 2(c), respectively. In thicker layers, additional micron-scale surface corrugations oriented parallel to the striations become evident, as shown in the large-area scan of sample F in Figure 2(d). Similar surface features spanning multiple length scales have been reported previously[19,22]. Our observations indicate that both their origin and orientation are directly linked to the elongated surface features formed during the pre-growth annealing step. The root-mean-square (RMS) roughness as a function of epilayer thickness is summarised in Figure 2(e). Among the samples studied, the lowest RMS roughness is observed for the 1.2 µm-thick sample C (2.86 ± 0.10 nm). At a thickness of 0.3 µm, the roughness is higher (4.17 ± 0.51 nm), likely due to incomplete coalescence. For thicknesses exceeding 1.2 µm, no further post-coalescence smoothing is observed. Instead, the surface roughness increases monotonically, reaching (5.46 ± 0.95) nm at a thickness of 3.5 µm for sample F. Importantly, twinning, which is common for MOCVD growth on (100)-oriented gallium oxide[25,26], was not observed.

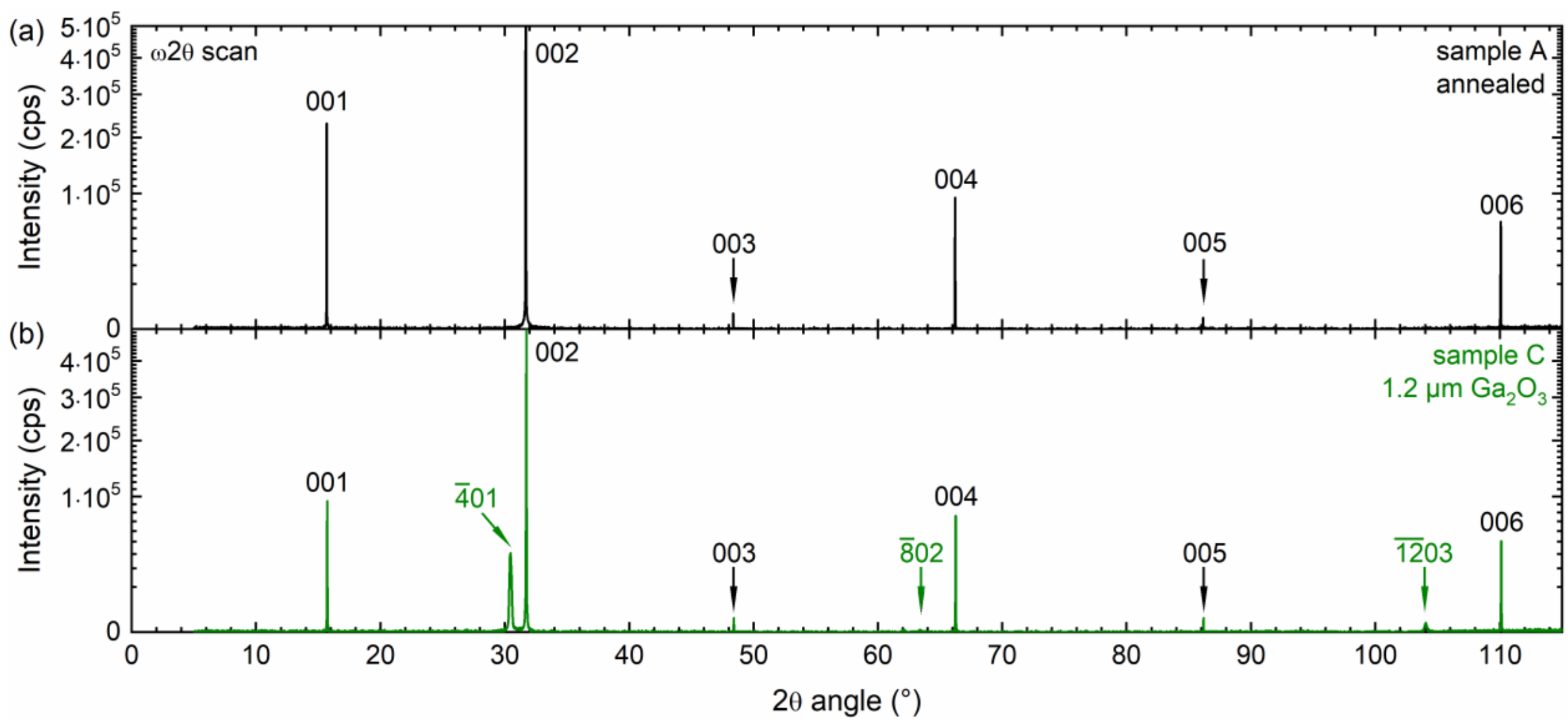

*Figure 3: HRXRD ω–2θ scans of (a) the annealed β-$Ga_2O_3$ substrate (sample A) and (b) the sample with 1.2 µm thick epilayer (Sample C).*

Next, we investigated the crystalline structure of the epilayers using double-axis high-resolution X-ray diffraction (HRXRD). These measurements were performed to evaluate the crystal orientation of the as-grown epilayers in relation to the underlying substrate. For comparison, Figures 3(a) and 3(b) show large-range ω-2θ scans for the annealed sample A and for a representative grown sample, sample C, with the 1.2 µm thick epilayer, respectively. For sample A, the diffractogram exhibits the 001 reflection of β-$Ga_2O_3$ and its higher-order harmonics, consistent with the (001) surface orientation of the substrate. In contrast, Figure 3(b) reveals additional prominent peaks corresponding to the -401, -802, and -1203 reflections of β-$Ga_2O_3$. These peaks were present in all the samples B-F (see Figure S1 in the Supporting Information).

Previous studies have also reported -401 reflections. For example, Meng *et al.* identified a -401 peak in both 1.5 µm and 9 µm thick epilayers grown on (001) substrates[22], while Qin *et al.* reported similar -401 peaks in all their samples grown over a range of growth conditions[21]. Furthermore, close inspection of the HRXRD data in other reports reveals unlabelled but intense peaks at approximately 2θ ≈ 31° and 64°, consistent with the expected positions of the -401 and -802 reflections, respectively[24]. These observations underscore the generality of the present findings.

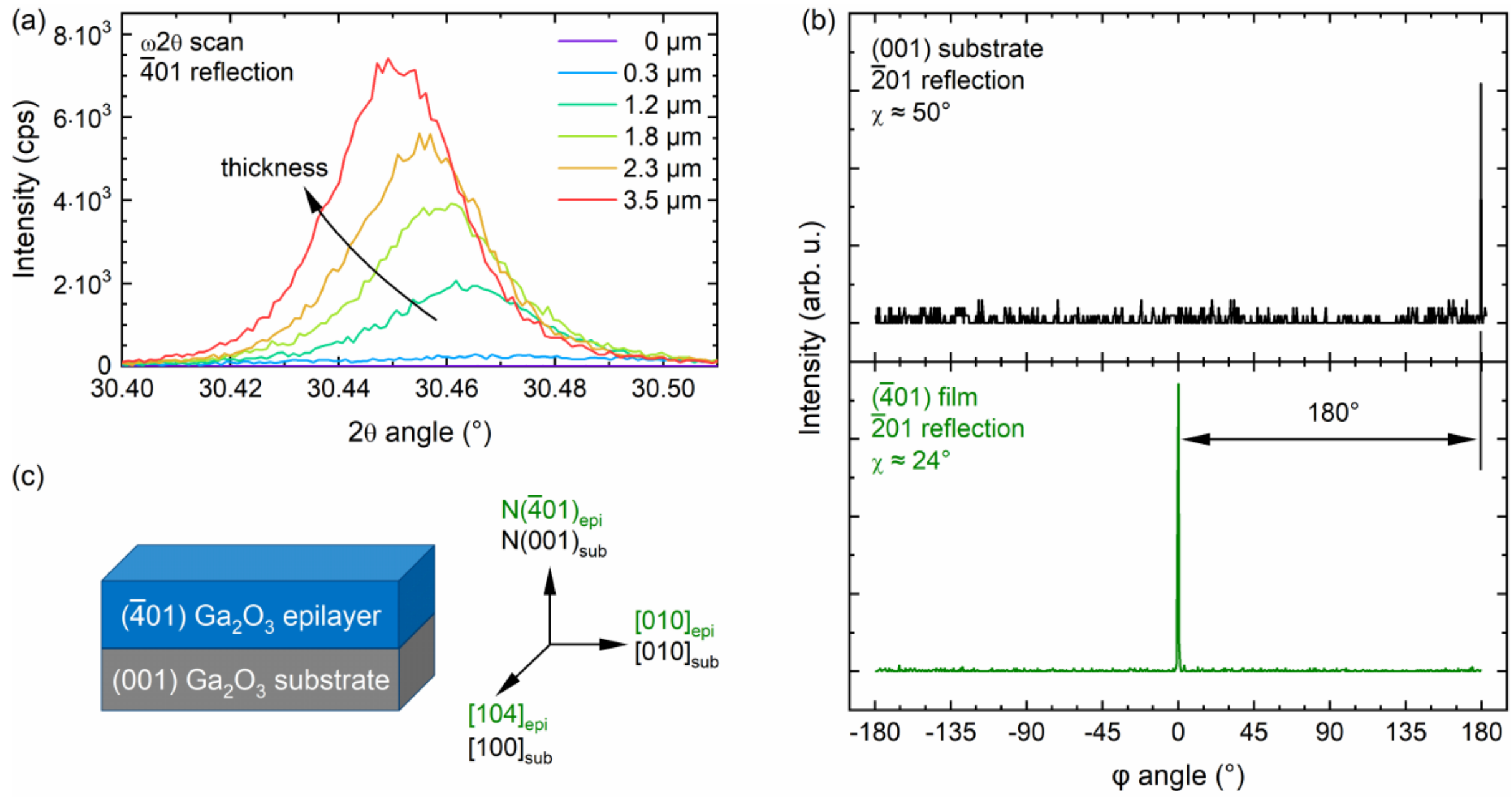

*Figure 4: (a) HRXRD ω–2θ scan, showing the evolution of the (-401) reflection with increasing thickness of the β-$Ga_2O_3$ epilayer. (b) Azimuthal (φ) scans of the -201 reflections of the (001) substrate and the (-401) epilayer to determine the in-plane orientation illustrated by the schematic in (c).*

Further analysis of the -401 peak in our samples shows that its intensity increases with epilayer thickness, while its peak width decreases, as shown in Figure 4(a). A systematic shift of the -401 peak towards lower diffraction angles with increasing epilayer thickness is also observed, indicative of a progressive evolution in the lattice parameters, as discussed later. In contrast, no such systematic behaviour was observed for the 001 reflection. Furthermore, a wide 002 peak superimposed on the narrow substrate peak expected for an (001) epilayer is absent (see Figure S2 in the Supporting Information). These results provide strong evidence that the detectable epilayer volume is dominated by a single (-401)-oriented domain along the growth direction, rather than consisting of (-401) rotational subdomains embedded within an otherwise (001)-oriented film, as previously assumed by others[21,22].

To establish the crystallographic relationship between the (-401)-oriented epilayer and the (001) substrate, azimuthal (φ) scans were performed in skew-symmetric geometry ($\chi \neq 0°$). Figure 4(b) shows such a representative scan for the well-distinct off-axis -201 reflection of both the (-401) epilayer and the (001) substrate. In each scan, only one -201 reflection is present, separated by 180° between the substrate peak and epilayer peak. This indicates that the epilayer grows in a single orientation without in-plane twinning and that the in-plane epitaxial relationship between the (-401) epilayer and (001) substrate is

$$[010]_{epi} \parallel [010]_{sub} \text{ and } [104]_{epi} \parallel [100]_{sub},$$

as indicated by the sketch in Figure 4(c). Consequently, the striations and corrugations observed in Figures 2(a)–2(d) are aligned parallel and perpendicular to the epilayer [010] and [104] directions, respectively. In this sense, the growth—while chemically homoepitaxial—deviates from ideal crystallographic homoepitaxy.

The full width at half maximum (FWHM) of HRXRD ω-rocking curves is commonly used as a metric of epilayer mosaicity. Here, the broadening of on-axis reflections is primarily influenced by mosaic tilt and strain in the specific measurement direction, whereas off-axis reflections measured at a large skew angle are sensitive mainly to in-plane twist. Figure 5 summarises the FWHM values obtained from both on-axis and off-axis reflections. For the (-401)-oriented epilayers, the ω-FWHM of the off-axis 002 reflection ($\chi \approx 69.6°$) is significantly larger than that of the on-axis -401 reflection, suggesting a stronger twist than tilt within all epilayers. Notably, the FWHM values of the on-axis -401 reflections measured along both in-plane directions are comparable for all samples except the thinnest 300 nm epilayer. For this sample, the reflection measured co-planar with the $[104]_{epi}$ direction exhibits a markedly smaller FWHM. The origin of this behaviour is discussed later. For epilayer thicknesses ranging from 1.2 µm to 3.5 µm, the ω-FWHM values of all reflections decrease progressively, indicating a reduction in tilt and twist of individual mosaic domains with increasing thickness.

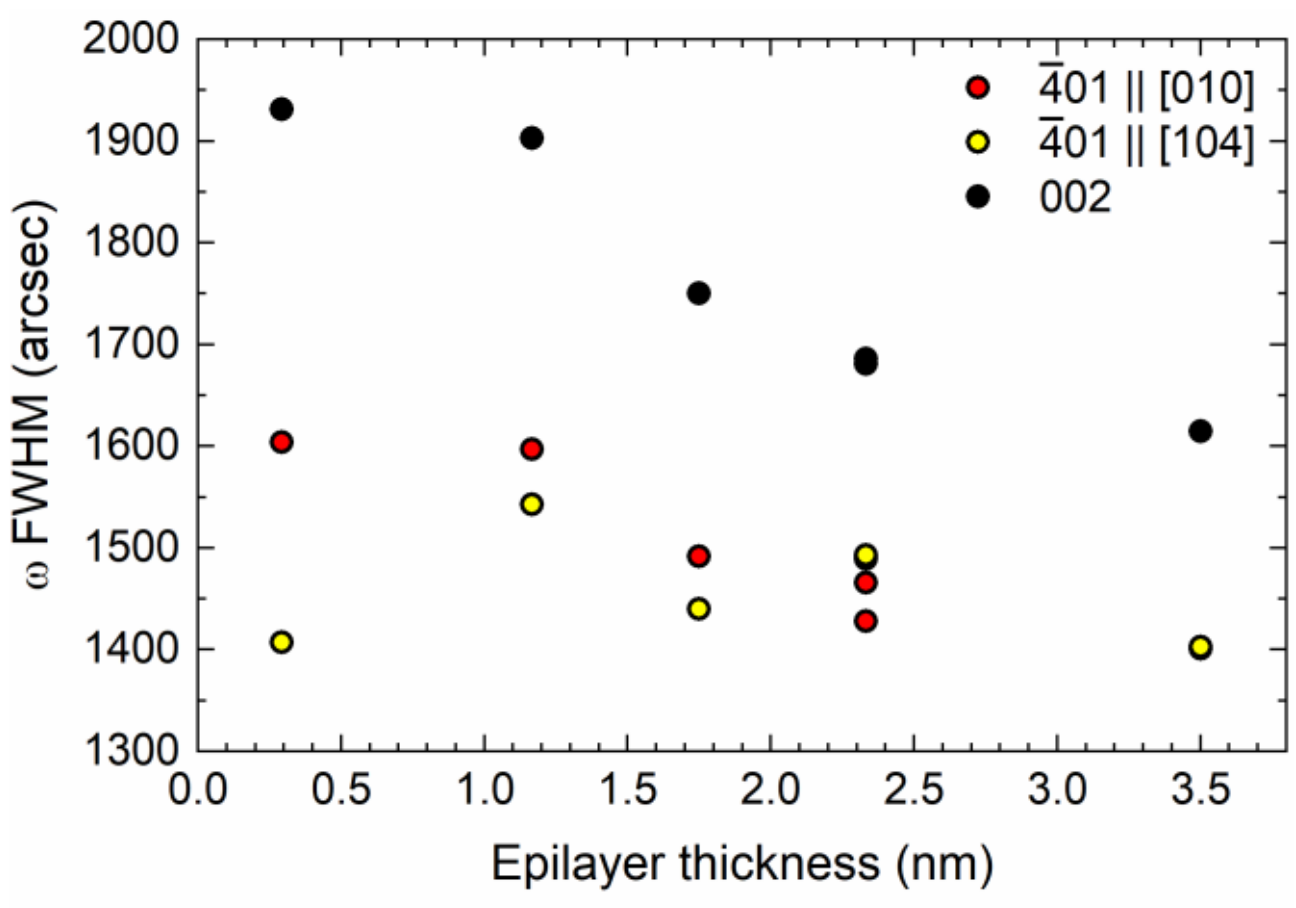

*Figure 5: Evolution of the ω-FWHM values for the -401 and 002 reflections for the (-401) β-$Ga_2O_3$ epilayers with thickness. The on-axis -401 reflection was measured in symmetric geometry along the two in-plane directions [010] and [104], whereas the off-axis reflection was measured in pure skew-symmetry.*

Figure 6 shows representative large-area optical micrographs of the samples. All our MOCVD-grown epilayers exhibit mirror-smooth surfaces at the macroscopic scale. However, channelling cracks, which extend from one edge of the sample to the other, appear at the epilayer thickness of ~1.8 µm (Figure 6(c)) and persist at greater thicknesses (Figure 6(d)). Notably, this critical thickness is comparable to that reported by Meng *et al.*[22]. The cracks in our samples are consistently oriented parallel to the epilayer [010] direction. Although a quantitative analysis was not undertaken, a comparison of the large-area optical micrographs of Figure 6(c) and 6(d) clearly shows an increase in crack density with epilayer thickness. The observed thickness-dependent onset of cracking suggests a strain-driven process. We briefly discuss this next, considering the framework of Griffith's fracture criterion, which states that crack formation becomes energetically favourable when the reduction in elastic strain energy exceeds the increase associated with the creation of new surface area, thereby minimising the total free energy of the system[27].

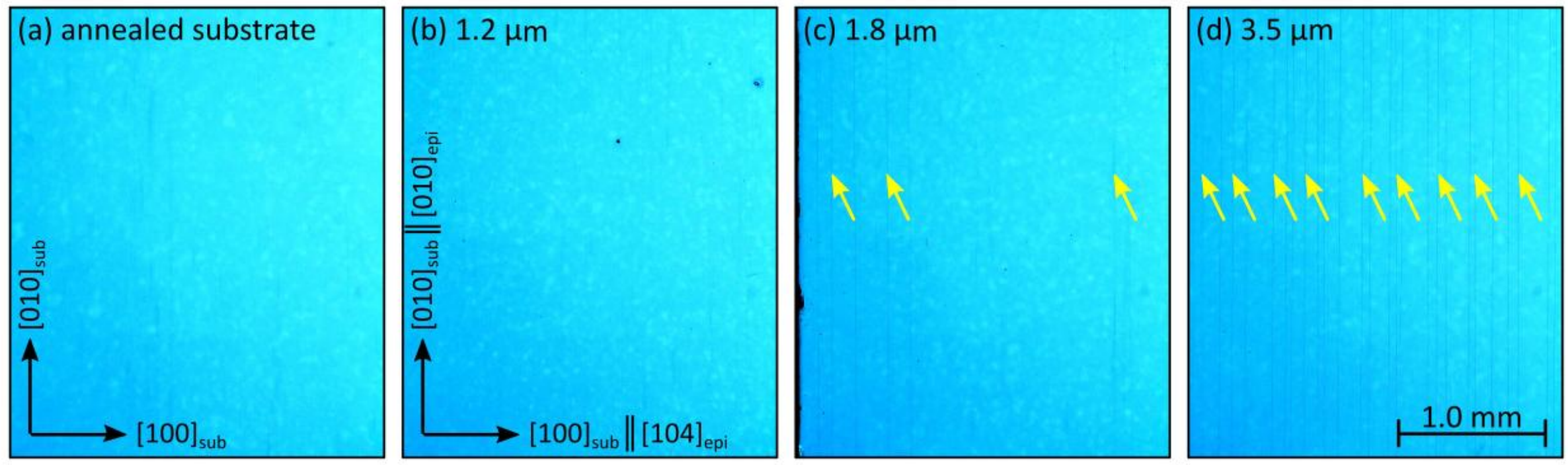

*Figure 6: Representative large-area optical microscopy images showing the β-$Ga_2O_3$ surface morphologies of (a) the annealed substrate and samples with epilayer thicknesses of (b) 1.2 µm, (c) 1.8 µm and (d) 3.5 µm. Channelling cracks become apparent at increased thickness, with a higher crack density observed in (d) compared to (c). For the specific fields of view shown, approximately six and twenty-seven cracks can be distinguished in (c) and (d), respectively. Arrows in (c) and (d) indicate individual cracks.*

Gallium oxide possesses poor thermal conductivity (~13 W $m^{-1}$ $K^{-1}$ along [001][28]), and a vertical thermal gradient—and therefore thermal strain—can be expected in the substrate during high-temperature growth. However, the absence of cracking in the annealed control sample and samples B-C suggests that thermally induced strain is unlikely to be the dominant origin of cracking. Also, as the substrates are diced before growth, wafer-scale thermal stress accumulation is further minimised, supporting this interpretation. We therefore consider anisotropic lattice misfit strain between the epilayer and substrate. Along the [010] direction, $Ga_2O_3$ epilayer and substrate have the same periodicity, and hence no lattice mismatch strain ($\varepsilon_{[010]} = 0$). In contrast, alignment of the epilayer [104] direction with the substrate [100] direction produces a large theoretical tensile in-plane strain in the epilayer of approximately +4.1%. This relation arises from the epitaxial alignment $[104]_{epi} \parallel [100]_{sub}$, for which two substrate lattice spacings along [100] are required to approximately match a single epilayer periodicity along [104] ( $L_{[104]}$ = 23.464 Å, $L_{[100]}$ = 12.210 Å, $\varepsilon_{[104]} = (2 \times L_{[100]} - L_{[104]}) / L_{[104]}$ where $L_{[uvw]}$ is the lattice periodicity in the direction [uvw]). This value represents the nominal geometric lattice mismatch strain; the actual elastic strain in the epilayer is expected to be partially relaxed through defect formation and crack propagation, as discussed below.

The observed crack orientations (⊥ to the epilayer's [104] direction) provide compelling evidence that this tensile misfit strain is the dominant driving force for crack formation. Supporting this conclusion, Figure 4(a) shows that the -401 XRD peak shifts systematically toward lower diffraction angles with increasing epilayer thickness. Based on the Bragg relation, $\lambda = 2d_{(-401)} \sin\theta$, this shift implies an increase in the out-of-plane (-401) lattice spacing and, correspondingly, a decrease in the in-plane [104] lattice constant, consistent with progressive relaxation of tensile in-plane strain as the epilayer thickness increases. It is also noteworthy that, in thicker samples, some cracks visible under optical microscopy did not exhibit surface opening in AFM scans of the same regions. This observation suggests that cracks may form at different stages during growth, with some subsequently overgrown. Such behaviour also supports the conclusion that cracking is driven primarily by epilayer–substrate mismatch strain rather than by differences in thermal expansion coefficients between the epilayer and substrate. If thermal mismatch were dominant, cracking would be expected to occur predominantly during post-growth cooling, as is commonly observed in other material systems, for example, in GaN grown on silicon substrates by MOCVD[29,30].

Further evidence of strain accumulation and incomplete relaxation may be inferred from the evolution of the surface morphology. Surface roughening is often symptomatic of strained growth in hetero-epitaxy[31,32], and the observed monotonic increase in roughness beyond coalescence in our samples may reflect the persistence of residual strain in the growing homo-epitaxial epilayers. In addition, growth cusps formed during epitaxy can act as stress concentrators[31]. Given that the surface corrugations in our samples are also oriented along $[010]_{epi}$, as inferred from the epitaxial relationship shown in Figure 4(c) and the morphology of Figure 2(d), these cusps may facilitate crack propagation along them. It is worth noting that, because the samples were diced using a saw before growth, dicing-induced damage at the sample edge was occasionally observed to act as a nucleation site for crack initiation (see Figure S3 in the Supporting Information).

Overall, our observations indicate that crack opening is the primary mechanism for releasing tensile strain energy in the thicker epilayers. Nonetheless, in addition to surface roughening, other mechanisms — such as the formation and propagation of extended defects including dislocations — may also contribute to strain relaxation from an early stage, as commonly observed in lattice-mismatched heteroepitaxial systems[33–35]. The increased ω-FWHM of the -401 reflection measured coplanar to $[104]_{epi}$ (Figure 5) for the 1.2 μm thick sample, compared with the 0.3 μm sample where cracks are absent, suggests that such defect-mediated relaxation mechanisms may become active at intermediate thicknesses. Confirming their role and identifying the operative glide systems would require detailed electron microscopy studies over statistically significant areas, which are beyond the scope of the present work.

Finally, we discuss the origin of the emergence of the (-401) orientation on nominally (001)-oriented substrates. The surface energies of $\beta$-$Ga_2O_3$ are highly anisotropic, with the (001) surface exhibiting one of the highest reported surface energies (1.95 J $m^{-2}$ for (001)A)[36]. The propensity of this surface to reduce its energy via structural modification has been demonstrated by *in situ* diffraction studies during molecular beam epitaxy[37]. The study showed that oxygen plasma exposure at ~950 °C induces reconstruction of the (001) surface along the [010] azimuth, yielding a surface which remains stable with no further reconstruction occurring during cooldown[37].

Although the MOCVD environment precludes equivalent real-time diffraction probes, the high oxygen chemical potential present during both annealing and growth is expected to drive similar surface modification. Supporting this interpretation, Figure 7(a) shows a high-resolution (300 × 300 $nm^2$) AFM image of the post-anneal control sample (sample A), revealing that facets elongated along the $[010]_{sub}$ direction have periodicities on the order of several tens of nanometres along $[100]_{sub}$. The close correspondence between this periodicity and the previously reported jagged epilayer–(001) substrate interfaces[22] is consistent with the interpretation that the surfaces shown in Figures 1(b) and 7(a) correspond to the faceted substrate surface present immediately before MOCVD growth. These facets are expected to correspond to low-index planes with small inclination angles relative to (001).

A schematic side-view of this proposed faceted reconstruction is shown in Figure 7(b). We hypothesise that under the non-equilibrium, oxygen-rich conditions characteristic of MOCVD, this reconstructed and faceted (001) substrate surface represents a low-energy configuration that may promote β-$Ga_2O_3$ growth with the (-401) planes parallel to the epilayer surface. Figure 7(b) illustrates the crystallographic relation between epilayer and substrate, which is achieved by rotating the epilayer unit cell by approximately 74° around the [010] zone axis with respect to the substrate. This inclination finds its origin in the relative orientation of the (-401) and (001) planes in the monoclinic unit cell, shown in Figure 7(c).

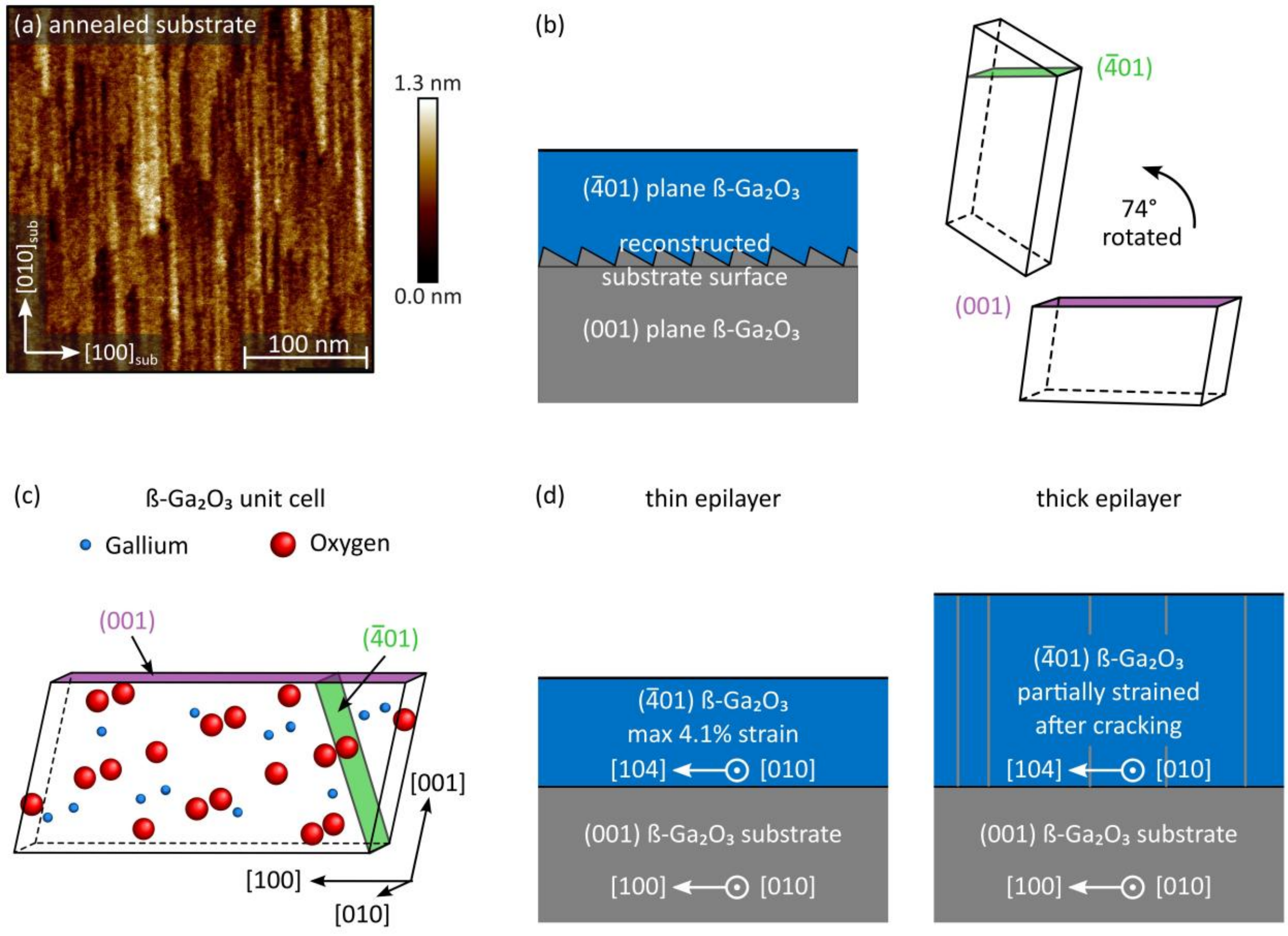

*Figure 7. (a) Representative high-resolution (300 × 300 nm²) AFM image of the annealed (001) β-$Ga_2O_3$ substrate (control sample A), showing elongated surface facets along [010]. (b) Schematic illustration of the (-401) β-$Ga_2O_3$ growth on (001) β-$Ga_2O_3$, highlighting the crystallographic orientation relationship between the β-$Ga_2O_3$ unit cells of epilayer and substrate; and (c) β-$Ga_2O_3$ unit cell illustrating the relative orientation of the (-401) and (001) crystal planes. (d) Schematic representation of the resulting strain state in the epilayer (left) arising from this orientation relationship, and its relaxation via crack formation with increasing epilayer thickness (right).*

Consequently,  the ensuing in-plane epitaxial alignment ($[010]_{epi}$ ∥ $[010]_{sub}$ and $[104]_{epi}$ ∥ $[100]_{sub}$) essentially introduces no lattice mismatch along the [010] direction, but produces a large tensile mismatch along the epilayer [104] direction (Figure 7(d)), which ultimately leads to cracking as the epilayer thickens to relieve the tensile strain energy. It is worth noting that our growths performed without the annealing step yielded identical results, indicating that surface reconstruction likely occurs at conventional growth temperatures and oxygen partial pressures even in the absence of deliberate annealing.

Lastly, we note that the use of $(Al_xGa_{1-x})_2O_3$ buffer layers has been demonstrated to increase the achievable thickness of crack-free $Ga_2O_3$ films[23], owing to the introduction of compressive strain associated with the smaller lattice constant of the $(Al_xGa_{1-x})_2O_3$. While this approach is certainly beneficial for the realization of thicker top layers, the large conduction band offset at the $(Al_xGa_{1-x})_2O_3/Ga_2O_3$ heterointerface is expected to impede vertical electron transport, particularly in device architectures where the doped substrate serves as the electrical contact. In contrast, strategies aimed at suppressing surface reconstruction and stabilizing epitaxial growth in the (001) orientation—such as temperature ramping under controlled gas ambients or the use of low-temperature $Ga_2O_3$ nucleation layers—may offer a more broadly applicable pathway toward achieving thick, high-quality layers.

## Conclusion

In summary, we have carried out a systematic investigation of MOCVD-grown $\beta$-$Ga_2O_3$ epilayers on nominally (001)-oriented native substrates, focusing on the evolution of surface morphology, crystal orientation, and cracking behaviour with epilayer thickness. Despite growth on (001) substrates, HRXRD analysis reveals that the epilayers adopt a predominantly (−401)-oriented structure from the earliest stages of growth. This orientation is closely linked to the reconstructed and faceted substrate surface formed under oxygen-rich MOCVD conditions and leads to a large tensile in-plane strain arising from lattice mismatch along the epilayer $[104]_{epi}$ in-plane direction, while it is perfectly lattice matched along $[010]_{epi}$. As the epilayer thickness increases, this anisotropic mismatch-driven strain progressively relaxes, ultimately resulting in crack formation parallel to the $[010]_{epi}$ direction. Our results demonstrate that the unique epitaxial relationship between the (-401)-oriented epilayer and the (001) substrate fundamentally limits the achievable crack-free thickness under conventional MOCVD conditions. These findings highlight the need for alternative nucleation strategies or growth-condition engineering to stabilise (001)-oriented epitaxy for device-relevant thick layers.

---

## Materials and Methods

### Growth

All growth experiments were carried out in an AIXTRON close-coupled showerhead (CCS) MOCVD reactor configured for substrates up to 100 mm in diameter. The system is equipped with an emissivity-corrected pyrometer (LayTec EpiTT) and an *in situ* X-point reflectometry system. Square 13.5 × 13.5 mm² β-$Ga_2O_3$ samples were used as substrates for growth. These were diced from an Sn-doped n-type β-$Ga_2O_3$ wafer with an estimated carrier concentration of $1.2 \times 10^{19}$ $cm^{-3}$, a thickness of 650 μm, and a diameter of 150 mm. The wafer was manufactured by Tamura Corporation using the EFG method and possessed a ~0.2° miscut from the surface (001) plane. Before reactor loading, the diced substrates were cleaned sequentially in organic solvents to remove organic contaminants, followed by an unheated sulfuric acid–hydrogen peroxide mixture to eliminate metallic and inorganic residues. Reported temperatures correspond to the pyrometer-measured temperature of the wafer pockets. TMGa (6N purity) and oxygen (5N purity) supplied from a gas cylinder served as Ga and O precursors, respectively, with nitrogen used as the carrier gas. The investigated $Ga_2O_3$ epilayers were not intentionally doped, but the reported growth rates were determined from separate calibration runs using silicon-doped epilayers grown under otherwise identical conditions, with secondary ion mass spectrometry (SIMS) used to identify the Si marker layer and determine epilayer thickness.

**Material Characterisation**

Surface topography measurements were performed using a Bruker Dimension Icon XR atomic force microscope operated in PeakForce tapping mode with ScanAsyst Air probes (nominal tip radius ~ 2 nm). All AFM images were flattened using a two-dimensional background subtraction using Nanoscope Analysis software. X-ray diffraction measurements were carried out on a Philips X'Pert Pro diffractometer using Cu $K\alpha_1$ radiation to irradiate ~ 1.5–2 $cm^2$ of the sample area and a gas-proportional point detector. Wide range ω-2θ scans and ω-rocking curves were collected in double-axis geometry, while short ω-2θ scans of -401 reflection were measured in triple-axis geometry. Optical micrographs were acquired using a Keyence VHX-950F optical microscope.

**Acknowledgements**

The authors acknowledge funding for 'High Voltage Vertical Diodes on Homoepitaxial Gallium Oxide' project, awarded by the Henry Royce Institute for Advanced Materials through the Industrial Collaboration Programme, funded from a grant provided by the Engineering and Physical Sciences Research Council (EPSRC), EP/X527257/1. The authors also acknowledge the support of Dr Andrew Pakes (AIXTRON) in facilitating this collaboration. The MOCVD reactor used for all epitaxial growth was funded through EPSRC Strategic Equipment funding, EP/T019085/1. S. Ghosh further acknowledges a SACEME Seedcorn grant from the Faculty of Science and Engineering, Swansea University, which supported the HRXRD measurements.

**Data Availability**

The data that support the findings of this study are available from the corresponding author upon reasonable request.

# Supporting Information

## Why Do Thick MOCVD-Grown β-$Ga_2O_3$ Epilayers on (001) Substrates Crack: Crystallographic Origin

Martin Frentrup[1,*], Indraneel Sanyal[2,*], Dan Lamb[3], Ciaran P. Llewelyn[3], Jonathan Evans[3], Owen J. Guy[3,4], Mike Jennings[3,5], Saptarsi Ghosh[3,5,#]

[1]Department of Materials Science and Metallurgy, University of Cambridge, Cambridge, CB3 0FS, UK
[2]AIXTRON Ltd, Buckingway Business Park, Anderson Rd, Swavesey, Cambridge CB24 4FQ, UK
[3]Centre for Integrative Semiconductor Materials, Swansea University, Swansea SA1 8EN, UK
[4]Department of Chemistry, Swansea University, SA1 8EN, UK
[5]Department of Electronic and Electrical Engineering, Swansea University, SA1 8EN, UK

*These authors contributed equally.

# Author to whom correspondence should be addressed: saptarsi.ghosh@swansea.ac.uk

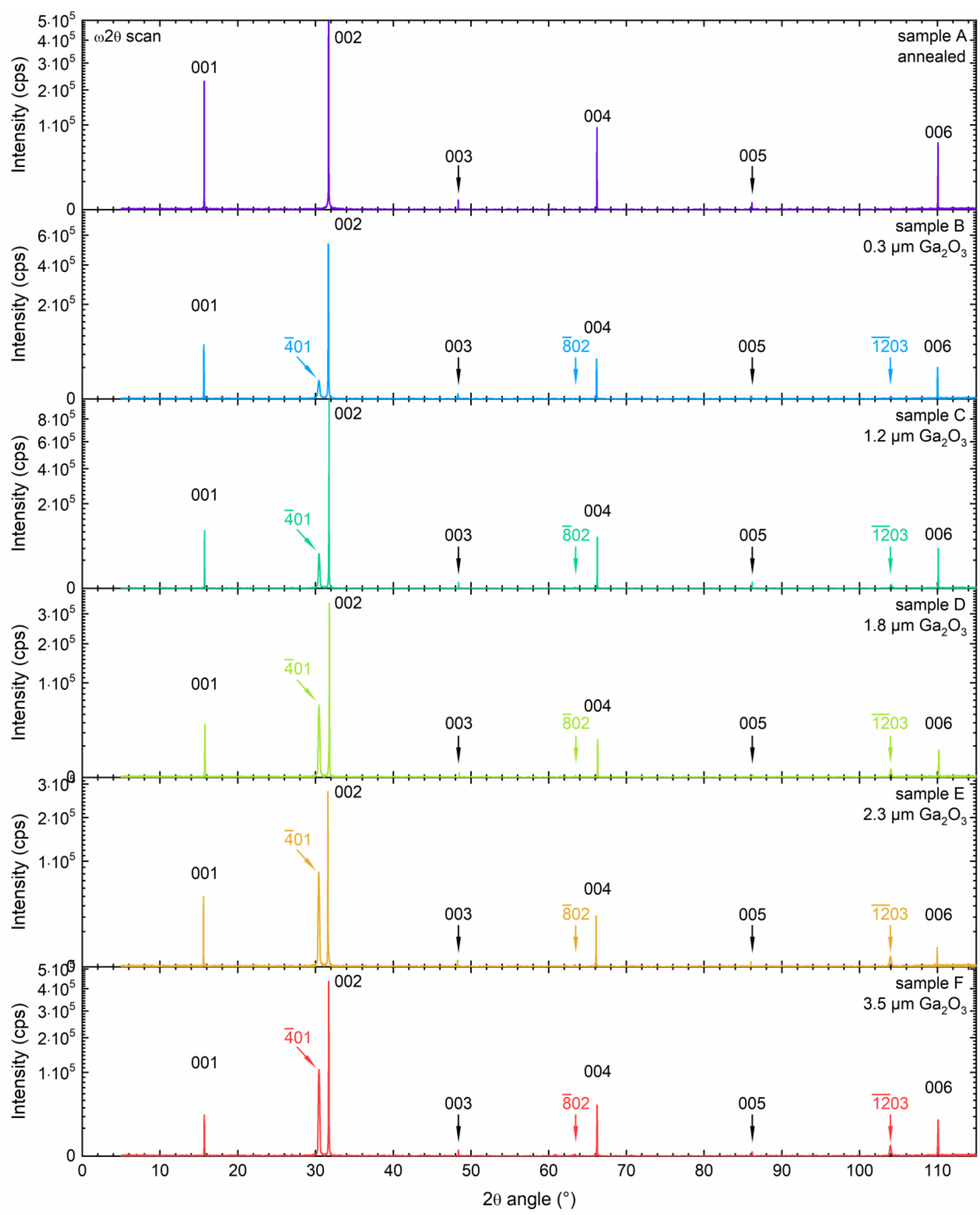

*Figure S1. HRXRD ω–2θ scans of the samples A-F, showing the emergence and evolution of the (-401) epilayer reflections with increasing epilayer thickness.*

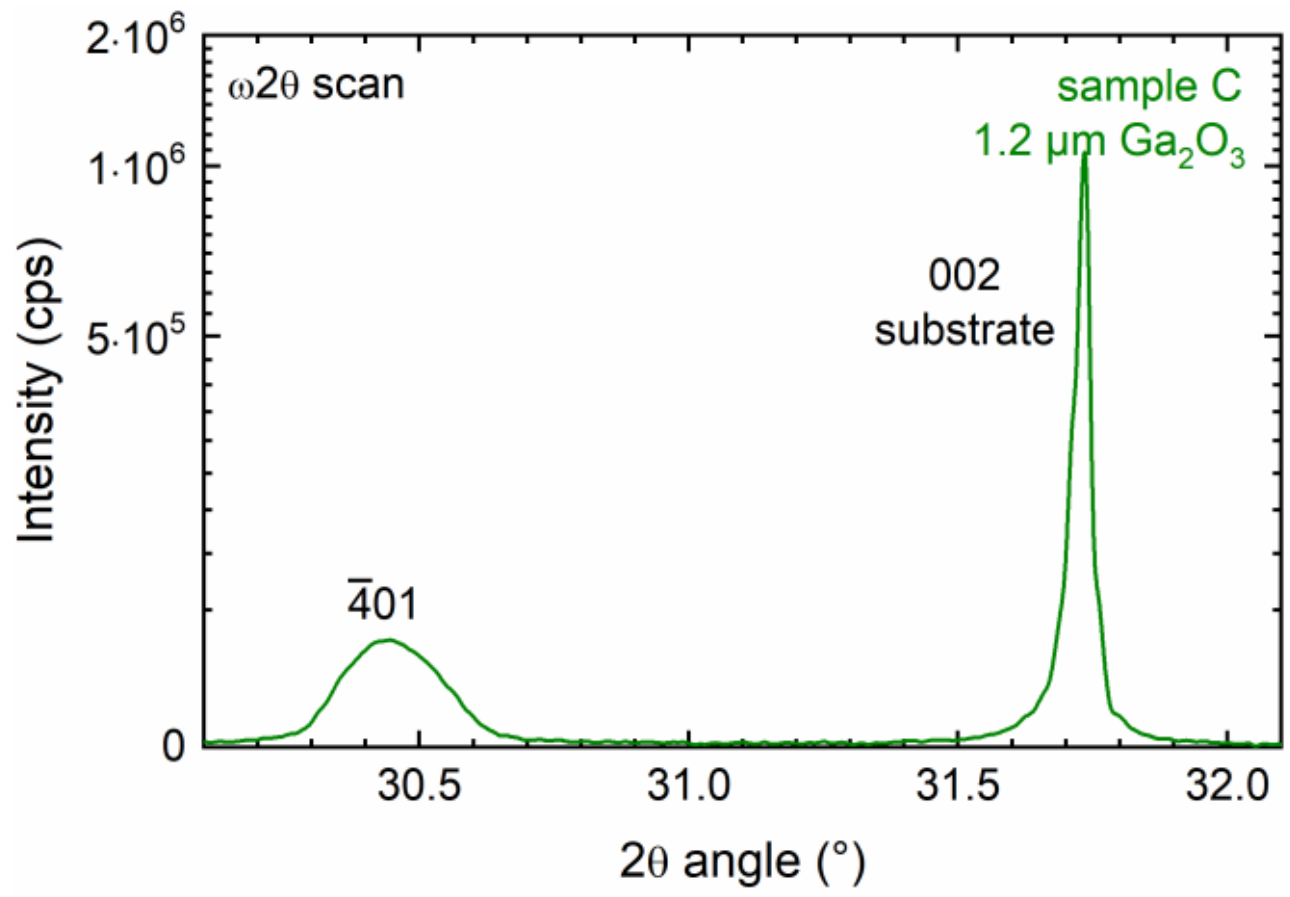

*Figure S2. Representative HRXRD ω–2θ scans of the sample with 1.2 µm thick epilayer (Sample C), showing the -401 reflection of the epilayer and the 002 reflection of the substrate. The absence of a second wide 002 substrate peak superimposed on the narrow substrate peak confirms that the epilayer does not grow in the (002) orientation, but instead grows as a single (-401) oriented film.*

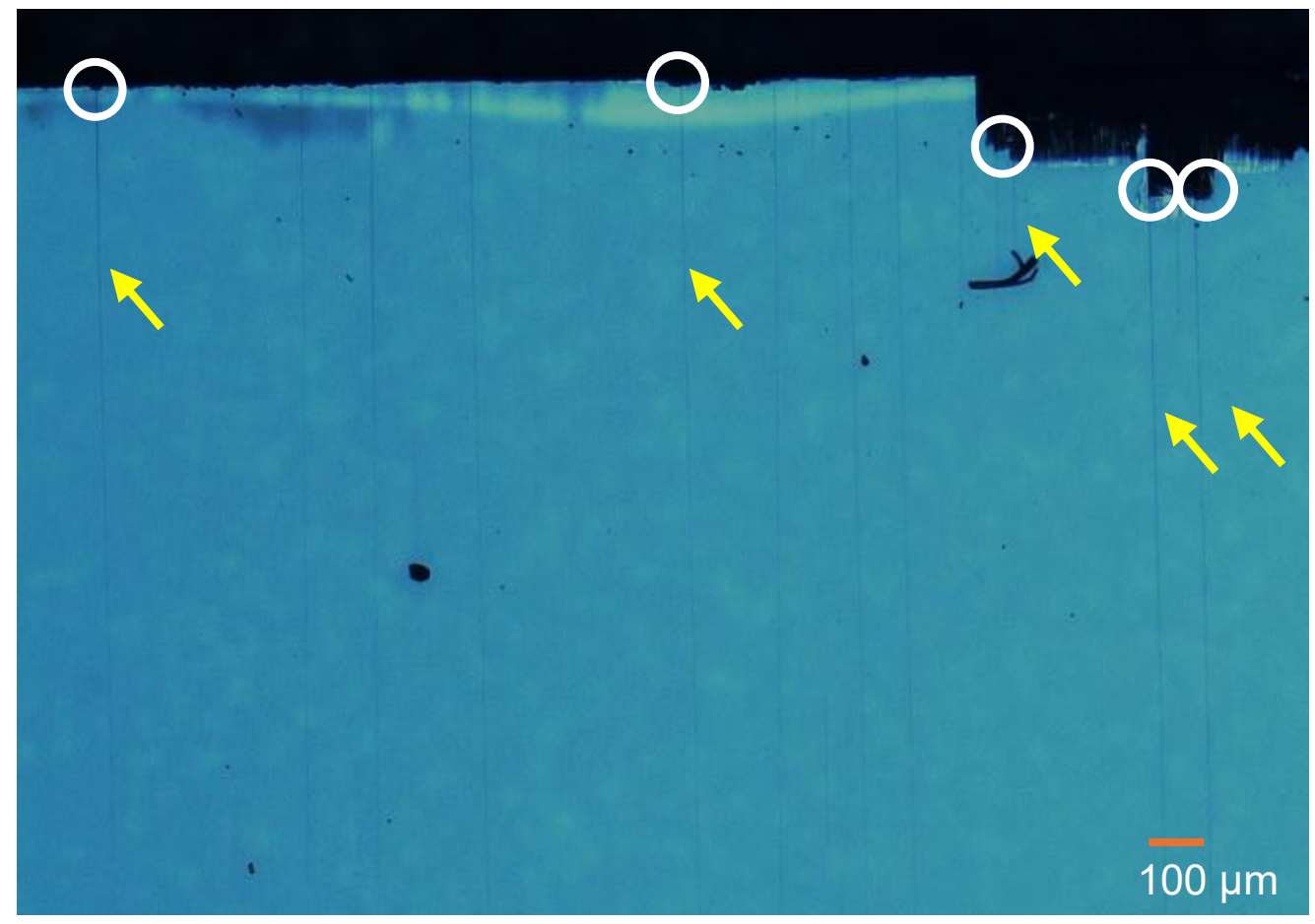

*Figure S3. Representative large-area optical microscopy image near the edge of a sample with an epilayer thickness of 1.8 µm. Due to dicing with a saw, the sample edges are not macroscopically abrupt and exhibit chipping-related damage, some of which are highlighted by white circles. Channelling cracks (indicated by a yellow arrow) are frequently observed to nucleate at these damaged regions. Although such edge non-uniformities are present in all samples, cracks are only observed in epilayers with thicknesses of 1.8 µm and above. This suggests that while these defects act as crack nucleation sites, a critical epitaxial strain is required for crack formation.*